\documentclass[10pt]{article}

\usepackage{amsmath}
\usepackage{amssymb}

\usepackage{graphicx}

\usepackage{cite}

\usepackage{color} 

\usepackage[labelfont=bf,labelsep=period,justification=raggedright]{caption}

\makeatletter
\renewcommand{\@biblabel}[1]{\quad#1.}
\makeatother

\date{}

\usepackage{lineno}

\renewcommand{\baselinestretch}{1.5}

\usepackage{bbold} 
\usepackage[table]{xcolor}

\usepackage{booktabs,caption,fixltx2e}
\usepackage[flushleft]{threeparttable}

\begin{document}
\thispagestyle{empty}
\noindent To appear in PLoS ONE

\begin{flushleft}
{\Large
\textbf{Timing of Pathogen Adaptation to a Multicomponent Treatment}
}
\\
Romain Bourget$^{1,2,3,4\ast}$, 
Lo\"ic Chaumont$^{1}$, 
Natalia Sapoukhina$^{2,3,4}$
\\
\bf{$^1$} Laboratoire Angevin de Recherche en Math\'ematiques, Universit\'e d'Angers, Angers, France
\\
\bf{$^2$} Institut National de la Recherche Agronomique, UMR1345 Institut de Recherche en Horticulture et Semences, Beaucouz\'e, France
\\
\bf{$^3$} AgroCampus-Ouest, UMR1345 Institut de Recherche en Horticulture et Semences, Angers, France
\\
\bf{$^4$} Universit\'e d'angers, UMR1345 Institut de Recherche en Horticulture et Semences, Angers, France

$\ast$ E-mail: bourget@math.univ-angers.fr
\end{flushleft}

\section*{Abstract}
The sustainable use of multicomponent treatments such as combination therapies, combination vaccines/chemicals, and plants carrying multigenic resistance requires an understanding of how their population-wide deployment affects the speed of the pathogen adaptation. Here, we develop a stochastic model describing the emergence of a mutant pathogen and its dynamics in a heterogeneous host population split into various types by the management strategy. Based on a multi-type Markov birth and death process, the model can be used to provide a basic understanding of how the life-cycle parameters of the pathogen population, and the controllable parameters of a management strategy affect the speed at which a pathogen adapts to a multicomponent treatment. Our results reveal the importance of coupling stochastic mutation and migration processes, and illustrate how their stochasticity can alter our view of the principles of managing pathogen adaptive dynamics at the population level. In particular, we identify the growth and migration rates that allow pathogens to adapt to a multicomponent treatment even if it is deployed on only small proportions of the host. In contrast to the accepted view, our model suggests that treatment durability should not systematically be identified with mutation cost. We show also that associating a multicomponent treatment with defeated monocomponent treatments can be more durable than associating it with intermediate treatments including only some of the components. We conclude that the explicit modelling of stochastic processes underlying evolutionary dynamics could help to elucidate the principles of the sustainable use of multicomponent treatments in population-wide management strategies intended to impede the evolution of harmful populations.\\
 

\section*{Introduction}

The emergence and spread of pathogen mutants able to overcome treatments that hitherto conferred complete protection on a host population, has become a real scourge in medicine, agriculture and forestry \cite{McD2002,Rea2007,Bos2008,Mac2010}. Multicomponent treatments, such as combination therapies simultaneously using several different antibiotics, recombinant multicomponent vaccines targeting more than one stage in the pathogen life cycle, mixtures of chemicals with differing mechanisms of action, and multigenic plant resistance carrying more than one resistance gene, were believed to be an efficient way to prolong the effectiveness of existing treatment components by delaying the pathogen adaptation process \cite{Lip1997,McD2002,Die2005,Hill2011}. Even though adapting to a multicomponent treatment involves multiple mutations, and therefore a higher cost to achieve adaptation, several phenomena, such as genetic drift, migration, recombination and the selective pressure exerted by the treatment itself, make it possible for an escape mutant to emerge. Striking examples such as the re-emergence of tuberculosis in a multidrug resistant form \cite{Blo1992}, seasonal adaptations of influenza to multicomponent vaccines \cite{Bon2008}, and the breakdown of multigenic plant resistance by foliar pathogens \cite{Mun1990}, have revealed that to achieve their purpose, multicomponent treatments should be deployed using optimal management strategies that control the adaptive dynamics of pathogens at the population level \cite{McD2002,Bos2008,Wie2011}.\\

On average, multicomponent treatments take fifteen to twenty years of investment, which makes them too expensive to be carelessly frittered by inappropriate use. Surprisingly, little attempt has been made to estimate the durability of multicomponent treatments employed in population-wide management strategies. This continuing ignorance can be explained by the fact that determining the speed at which a pathogen adapts at the population level requires considering processes as challenging to model as the stochastic emergence of an escape mutant and its spread throughout a host population, both of which can be altered by the management strategy. Most of the studies that have considered pathogen adaptation as a population-wide epidemiological problem use deterministic SIR compartment models in which disease transmission is modeled in terms of contacts between two types of host individuals, namely treated and untreated hosts \cite{Bon1997,Bos2003,Die2005,Gan2007,Cas2012}. The advantage of the approach is that it makes it possible to link the properties of the management strategy, for instance the spatial heterogeneity of the treatment \cite{Deb2009} or treatment coverage \cite{Cas2012,Fab2012}, with the dynamics of the pathogen. However, with rare exceptions \cite{Bon1997,Bos2003}, this approach is actually used to derive the invasion conditions of a pre-existing mutant, and not the speed of the pathogen adaptation. Another limitation is that compartmentalised models can track the dynamics of only limited numbers of types of host and pathogen \cite{Day2012}. Furthermore, treatment components can be used in various ways to devise a variety of population-wide control strategies that split the host population into several types. For instance, studying the impact of deploying two different antibiotics on bacterial evolutionary dynamics, Bonhoeffer \& al. \cite{Bon1997} demonstrated that treatment strategies in which, at any given time, equal fractions of the population receive different antibiotics can delay the trend toward antibiotic resistance. Management strategies intended to increase the environmental heterogeneity facing the pathogen, thereby inhibiting its spread, are becoming preponderant in the management of pathogen adaptive dynamics \cite{Bos2008,Wie2011}, since they offer the hope that the current arsenal of treatments can be used judiciously. However, recent theoretical studies have shown that the success of a deployed management strategy depends on the difference in relative fitness between the resident and mutant pathogen strains, which determines the intensity of their competition \cite{Ber2004,Deb2009,Mas1993,Gan2007}. This leads us to conclude that the evolutionary trajectory of a mutant escaping a multicomponent treatment included as part of a population-wide management strategy can be impressively complex, requiring the development of modelling approaches that account for both the life-cycle parameters of the different pathogen types, and the structure of the host population diversified by the management strategy. \\

Stochastic models are more appropriate for estimating the time it takes for evolutionary change to occur in a pathogen population spreading over heterogeneous environments, and could overcome limitations of the widely-used deterministic SIR approach \cite{Her2010}. Indeed, the interactions between stochastic migration and stochastic selection engender evolutionary processes that are invisible to deterministic evolutionary theory \cite{Ric2009}. However, the existing models are often just stochastic versions of SIR epidemic models, and even those that study pathogen adaptive dynamics at the population level ignore the stochastic migration process \cite{Res2006,Dag2007}. \\

The objective of this paper is to estimate the speed of pathogen adaptation in a host population which is subjected to a multicomponent treatment, and to provide some general guidance for the sustainable population-wide use of multicomponent treatments. To model the spread of the mutant through a host population, split by a management strategy into numerous types, we consider the mutant migration from an infected host, where it emerged, into a host receiving treatment. If the basic reproductive number of a new pathogen type is greater than one, then it is almost certain to become established, since there is no competition on the treated hosts. We therefore define {\it the emergence time} of an escape mutant as the time of the first migration from a mutant-infected host into a host subjected to multicomponent treatment. We formulated and analyzed a stochastic model based on a multi-type Markov birth and death process. This allowed us to take into account both pathogen mutation and migration stochastic processes, and the structure of a host population diversified by the management strategy. Unlike conventional studies of pathogen adaptive dynamics, our model allows us to track the population size dynamics of numerous pathogen types spreading through a heterogeneous host population. In numerical simulations, we varied the values of pathogen growth and migration rates, and determined the time to emergence of a mutant pathogen in a host receiving a multicomponent treatment. We first explored the impact of the proportion of hosts treated on the durability of the multicomponent treatment, and then looked at the impact of the mutation cost. We also investigated the effect of various management strategies on the durability of multicomponent treatment. We discuss the application of the results obtained to the population-wide deployment of combination therapies, combination vaccination/chemicals, and to the use of multigene resistance in cultivar mixtures.

\section*{Materials and Methods}

\subsection*{Model overview}

We assume that there are $N$ different treatment components that can be combined in various ways to devise a treatment for a host population. A treatment including all $N$ components is designated a multicomponent treatment, treatments consisting of only one component are designated monocomponent treatments, and other treatments, with from 2 to $N-1$ components, are designated intermediate treatments. Moreover, the application of these different treatment types can constitute a population-wide management strategy intended to diversify the host population when different numbers of host individuals receive one type of treatment only \cite{Bon1997}. Thus, as a result of the diversification resulting from the management strategy, the host population can contain up to $2^N$ types: one untreated and $2^N-1$ treated types all receiving different combinations of the treatment components. Since our model is non-spatial, it assumes that the management strategy applied will result in a well-mixed host population. \\

We consider pathogen population to be a set of infectious particles transmitting the disease, such as spores in fungal plant diseases, virus-laden aerosols or infectious agents carried by mosquitoes. An individual pathogen can be susceptible or resistant to any component of a $N$-component treatment. Thus, the pathogen population can contain up to $2^N$ pathogen types. The initial composition of the pathogen population includes a resident pathogen type produced by infected but untreated hosts, and can include pathogen types that have already adapted to monocomponent treatments. We start studying pathogen dynamics after the first stage of development of the epidemic, when all untreated hosts have already been infected, and the resident pathogen types have already reached a steady state with the host density. If we consider the deployment of a multicomponent treatment simultaneously with defeated treatments, we assume that all hosts carrying defeated treatments are also infected. In other words, the pathogen population starts with maximum population size for the resident pathogen type (and, if considered, adapted pathogen types) at time $t=0$. This initial state can be the result of the two following situations: (1) the deployment of a management strategy can precede the infection, such as the preventive vaccination of a part of the host population or the replacement of some susceptible hosts by resistant ones or (2) the strategy is deployed after the epidemic starts, such as chemical/drug treatments, vaccination or the deployment of genetically-resistant hosts. In the second case, for the sake of simplicity, we assume that the treatment immediately eliminates the infection. In both cases, it is assumed that only a part of the host population is treated.\\  

Let $X_{ij}(t)$ denote the random variable for the number of pathogens of type $i$ on host type $j$ at time $t\in\left[0,\infty\right)$, $(i,j) \in [1,2^N]^2$, with a state space consisting of the non-negative integers, $\left\{0,1,2,...\right\}$ . If pathogen type $i$ is not able to infect host type $j$, then for all $t, X_{ij}(t)=0$. The set $\left\{X_{ij}(t), t\in\left[0,\infty\right)\right\}$, $(i,j) \in [1,2^N]^2$, makes up the vector random variable $X(t) = \left(X_{11}(t),...,X_{12^{N}}(t),...,X_{2^{N}1}(t),...,X_{2^{N}2^{N}}(t)\right) $ representing the number of each type in the pathogen population at time $t$. Let $\left\{X(t),t\in\left[0,\infty\right)\right\}$ be a $d$-dimensional continuous-time, homogeneous, birth and death Markov process with state space $U=\left\{u=\left(u_1,...,u_d\right), u_j=0,1,...\infty, j=1,...,d\right\}$, where $d=2^{2N}$, of which the transition matrix only allows transitions to certain nearest neighbours \cite{Reu1961}. The model drives the dynamics of each pathogen type $i$ inhabiting host type $j$ by birth and death, mutation and migration stochastic processes. When a birth or death occurs, the population size of pathogen type, $X_{ij}(t)$, increases or decreases by one. Moreover, during its multiplication pathogen type $i$ can produce a mutant of type $k$, which is capable of infecting particular type(s) of treated hosts. At this moment the population size of pathogen type $k$ on host type $j$ increases by one. A given mutation is able to defeat only one component of the treatment. Since there is a mutation cost, and a high probability of being outcompeted by the resident type, the mutant's chance of surviving grows if it is transmitted to a suitable treated host, where there is no competition. We use the term 'migration' to refer to the transfer of an individual pathogen from one host to another; this can correspond to droplet-borne, sexually transmitted, or vector-borne disease transmission or the airborne transport of spores. Thus, when an individual pathogen of type $i$ migrates from host type $j$ to host type $l$, $l \ne j$, the population size of pathogen type $i$ on host type $j$ decreases by one, and that of $i$ on host type $l$ increases by one, if the host type $l$ is suitable and can be infected. Otherwise the migrants are lost and drop out of the pathogen dynamics. Note that migration does not reduce the severity of the disease on the host population, in a way that parallels the spread of spores in fungal diseases. The succession of stochastic birth, death, mutation and migration events provides a population-level description of the adaptation of a pathogen population to a treatment-diversified host population. Further, we define the transition probabilities for the above-listed stochastic events in a formal way.\\

\subsection*{Transition rates}

{\bf Mutation and birth transition rates.}\\
An individual pathogen can duplicate itself either to produce a new individual belonging to the same pathogen type or it can mutate into a new pathogen type. Since we assume that a pathogen mutant occurs during reproduction, a mutation event is considered as the spontaneous birth of an individual pathogen. The emergence of a individual pathogen $i$ as a result of mutation depends on the mutation probabilities of all the potential ancestors $k$ of type $i$. Then, the mutation transition rate is
\begin{equation}
\Lambda_{ij}(X(t)) = \sum_{k \ne i} p_{ki} r \beta_k X_{kj}(t),
\label{EqMut}
\end{equation}
where $p_{ki}$ is the probability that pathogen type $k$ will mutate into pathogen type $i$, $p_{ii}=0$, $r$ is the pathogen growth rate, and $\beta_k$ is the fractional reduction of pathogen type $k$ growth rate induced by additional mutation costs. We assume that the pathogen population undergoes one mutation per reproduction. Thus, if pathogen type $i$ can be attained only by means of several mutations, then $p_{ki}=0$, else $p_{ki}=\nu$. Thus, the probability that a mutation will produce pathogen type $i$ becomes $\sum_{k \ne i} p_{ki}$. We assume that successive mutations decrease the pathogen growth rate in an additive way \cite{Phi2008,Trin2009}. We define a cost $C_z$ ($z \in [1,N]$) for each mutation, such that the fractional reduction of pathogen type $k$ growth rate results in $\beta_k = 1-\sum_{z=1}^N C_z \mathbb{1}_{\left\{\mbox{The individual $k$ carries mutation $z$}\right\}}$, $\beta_k \in [0,1]$. Further, for the sake of simplicity, we assume that the mutation cost $C_z$ is the same for all mutations $z$, $C_z \equiv C$. Thus, pathogen types differ with regard to their fitness $r\beta_k, \; k \in [1,2^N]$.

We define the birth transition rate without mutation of pathogen type $i$ on host type $j$, $\lambda_{ij}(X(t))$, as a product between the growth rate of the pathogen type, $r$, its fractional reduction because of mutation cost, $\beta_i$, the pathogen type size, $X_{ij}(t)$, and the probability that no mutation will occur during a birth event, $1-p_i$, where $p_i= \sum_{l \ne i} p_{il}$ is the probability to mutate:
\begin{equation}
\lambda_{ij}(X(t)) =r \beta_i (1-p_i) X_{ij}(t).
\label{EqB}
\end{equation}

{\bf Migration transition rate.}\\
Since a migration event is the transfer of an individual pathogen from host type $j$ into host type $k$, we assume that its transition rate depends on the proportion of host type $k$ in the host population that has been diversified by a management strategy. Thus, the migration transition rate is,
\begin{equation}
\gamma_{ijk}(X(t)) = \xi_k D_{ik} X_{ij}(t),
\label{EqMigr}
\end{equation}
where $\xi_k$ is the proportion of host type $k$, in particular $\sum_{k=1}^{2^N} \xi_k = 1$, and $D_{ik}=D$, if pathogen type $i$ can infect the host $k$, or $D_{ik}=0$ if it cannot. Hereinafter, in numerical simulations, we use $D$, as the principle descriptor of the migration process, and we designate it the migration rate. When $N=1$, we have $\xi_1$ the proportion of untreated hosts and $\xi_2=1-\xi_1$ the proportion of treated hosts, but for the sake of simplicity, we use $\xi$ for $\xi_2$ and $1-\xi$ for $\xi_1$.\\

{\bf Death transition rate.}\\
We assume that the number of individual pathogens can decrease as a result of competition or by migration into unsuitable treated hosts. To include the negative effect of competition on the pathogen dynamics, we use the term,
\begin{equation*}
\frac{r \beta_i(1-p_i)X_{ij}(t)}{\xi_j K} \left( \sum \limits_{l=1}^{2^N} \frac{\beta_l }{\beta_i}X_{lj}(t)-1\right),
\end{equation*}
where $K$ is the maximum population size of the pathogen population; the maximum pathogen population size on host type $j$ is $\xi_j K$. The competitive relationships between $j$th and $l$th individuals are described by a generalized Lotka-Volterra equation for $2^N$ types. The competition intensity is determined by the ratio $\beta_l /\beta_i$. The term '$-1$' reflects the fact that an individual pathogen cannot compete with itself. Since a migration event is modeled by the death of an individual of pathogen type $i$ on host type $j$, and the birth of an individual pathogen of type $i$ on host type $k$, the mortality resulting from migration into unsuitable treated hosts is defined as follows, 
\begin{equation*}
X_{ij}(t)\sum \limits_{\substack{m=1,\\ m \ne i}}^{2^N}\xi_m(D- D_{im}),
\end{equation*}
where $D-D_{im}=0$ if pathogen type $i$ can infect the host $m$, or $D- D_{im}=D$ if it cannot. Then, the death transition rate of pathogen type $i$ on host type $j$ is 
\begin{equation}
\mu_{ij}(X(t)) = \frac{r \beta_i(1-p_i) X_{ij}(t)}{\xi_j K}\left( \sum \limits_{l=1}^{2^N} \frac{\beta_l }{\beta_i}X_{lj}(t)-1\right)+X_{ij}(t)\sum \limits_{ \substack{m=1,\\ m \ne i}}^{2^N}\xi_m(D- D_{im}).
\label{EqD}
\end{equation}

{\bf Interevent time.}\\
We define the probabilities of a birth, mutation, migration and death event as their transition rates divided by the sum of all transition rates. Following the definition in  \cite{All2003}, we assume the interevent time, the time between successive events, to be an exponential random variable with parameter:
\begin{equation}
\sum \limits_{ijk} \lambda_{ij}(X(t)) + \Lambda_{ij}(X(t)) + \gamma_{ijk}(X(t))+ \mu_{ij}(X(t)) \;.
\label{EqITime}
\end{equation}

\subsection*{Stochastic model of pathogen adaptive dynamics} 
The transition probabilities for the stochastic process $\left\{X(t),t\in\left[0,\infty\right)\right\}$, is $\mathbb{P}(X(t+\Delta t) = v | X(t) = u)$, $ u,v \in U$, where 

\[ X(t) = (\dots, X_{ij}(t), \dots, X_{ik}(t), \dots)=(\dots, a, \dots, b,\dots)=u\]
and
\[ X(t+\Delta t) = (\dots, X_{ij}(t+\Delta t), \dots, X_{ik}(t+\Delta t), \dots)=(\dots, a+m, \dots, b+n,\dots)=v\]
with $i,j,k \in [1,2^N]$ and $a,b,m,n \in \{1,2, \dots \}$, as follows
\begin{equation*}
\mathbb{P}(X(t+\Delta t) = v\, | \,X(t) = u)=
\end{equation*}
\begin{equation}\label{EqTransProb}
\left\{\begin{array}{ll}
(\lambda_{ij}+\Lambda_{ij})\Delta t+o(\Delta t), & \mbox{if} \;\; (m,n)=(1,0)\\
(\lambda_{ik}+\Lambda_{ik})\Delta t+o(\Delta t), & \mbox{if} \;\; (m,n)=(0,1)\\
\gamma_{ijk} \Delta t+o(\Delta t), & \mbox{if} \;\; (m,n)=(-1,1)\\
\gamma_{ikj} \Delta t+o(\Delta t), & \mbox{if} \;\; (m,n)=(1,-1)\\
\mu_{ij} \Delta t+o(\Delta t), & \mbox{if} \;\;  (m,n)=(-1,0)\\
\mu_{ik} \Delta t+o(\Delta t), & \mbox{if} \;\;  (m,n)=(0,-1)\\
1-(\lambda_{ij}+\Lambda_{ij}+\lambda_{ik}+\Lambda_{ik}+\gamma_{ikj}+\gamma_{ijk}+\mu_{ij}+\mu_{ik}) \Delta t+o(\Delta t), & \mbox{if} \;\; (m,n)=(0,0)\\
o(\Delta t), & \mbox{else}\end{array}
 \right.
\end{equation}
for $a+m \geq 0$, $b+n \geq 0$, and for $\Delta t$ sufficiently small. For the sake of the simplicity of notation in (\ref{EqTransProb}), we have set transition rates, $\lambda_{ij}(X(t))=\lambda_{ij}$, $\Lambda_{ij}(X(t))=\Lambda_{ij}$, $\gamma_{ijk}(X(t))=\gamma_{ijk}$, and $\mu_{ij}(X(t))=\mu_{ij}$, that are defined further by equations (\ref{EqB}-\ref{EqD}). We are interested in the emergence time of a mutant pathogen that occurs on a host receiving a multicomponent treatment: 
\begin{equation}\label{defS}
S=\inf\{t\ge0:X_{2^N2^N}(t) > 0\}\,,
\end{equation}
where $X_{2^N2^N}(t)$ is the population size of pathogen type with $N$ mutations able to invade hosts carrying an $N$-component treatment. Thus, the emergence time $S$ represents the time during which the multicomponent treatment is effective.

\subsection*{Model analysis}

In previous theoretical work, we analyzed the law of the emergence time of a particular mutant in a Markov chain \cite{Bou2013t}. In particular, we generalized conditions on the initial distribution under which the emergence time of a particular individual follows an exponential law. Here, to illustrate our theoretical findings, we analyzed numerically the distribution law of the emergence time in model (1-6). Simulations showed that, depending on the model parameters and settings, the distribution law of the emergence time can be closer either to an exponential or a normal law (Supporting Information, S1). If the host population includes individuals carrying intermediate treatments, the emergence time is well distributed around its mean as a normal law. In contrast, the absence of intermediate treatments prolongs the emergence time, resulting in an exponential law. The phenomenon intensifies with the increase in the number of treatment components, $N$. \\

Moreover, to gain a deeper understanding of how the model behaves, we carried out a global sensitivity analysis. The method used and results obtained can be found in the Supporting Information S2. Numerical simulations were then performed to identify the conditions under which the pathogen adaptation to a multicomponent treatment can be retarded. 

\subsection*{Simulation setup} \label{setup}

The adaptive dynamics of the pathogen population is modeled by tracking the population size of all pathogen types including the appearances and disappearances during stochastic birth and death processes. To perform numerical simulations, we used a range of  biologically-relevant parameter values corresponding to various diseases (Table \ref{param}). Sensitivity analysis showed that the effects of individual or joint variation of $K$ and $\nu$ on the pathogen dynamics are predictable, and that their increase reduces the emergence time. Thus, we fixed their values for the numerical simulations. We varied the values of the other life-cycle parameters of the pathogen population and properties of the management strategies. Each simulation was run until an escape mutant emerged on a host receiving the $N$-component treatment. In total, 1734 parameter sets were tested, and for each set, 1000 simulations were run to estimate the mean emergence time. The estimation of the confidence intervals of the emergence time means showed that they were significantly different. Since both the error bars and the confidence intervals were very small, they have been omitted in the figures presenting simulation results.\\  

{\bf The role of the proportion of the host treated in the pathogen adaptive dynamics.}\\
To study the behavior of the emergence time, we first estimated the waiting time for the first emergence of a mutant pathogen individual on a monocomponent treated host, $N=1$. At time $t=0$, the host population was split into two different types: untreated hosts and treated hosts, while the pathogen population at this time consisted of the resident type only. We varied the proportion of treated hosts  $\xi \in \{0.1,0.15,0.2,0.25,0.3,0.35,0.4,0.45,0.5,0.55,0.6,0.65,0.7,0.75,0.8\}$, the pathogen growth rate $r \in \{0.3, 0.4, 0.5, 0.6,1,2, 3, 6,12\}$, and the migration rate $D \in \{0.05,0.75,0.1,0.125,0.15,0.175,0.2,0.225,0.25,\\0.275,0.3\}$ (Fig. \ref{Fig1}). We fixed the mutation cost $C=0.2$, the maximum pathogen population size $K=10000$ and the mutation rate $\nu=10^{-5}$. The initial population size of the resident type on the untreated hosts was set to $(1-\xi) K$, while the sizes of the other types were zeroed. Then, we studied the emergence time behavior when the number of treatment components  was increased (not illustrated). We kept the management strategy and simulated pathogen adaptation for $N \in\{2,3\}$, five proportions of treated hosts, $\xi \in \{0.1,0.3,0.5,0.7,0.9\}$, the growth rate $r =  0.8$, the migration rate $D = 0.2$, the mutation cost $C=0.1$, the maximum pathogen population size $K=10000$, the mutation rate $\nu=10^{-3}$, and the same initial conditions. \\  

{\bf The effect of management strategies on the durability of the multicomponent treatment.}\\
To study the effect of the management strategy of $N$-component treatment on its durability, we varied the number of treatment components $N \in \{1,2,3,4\}$ for five distinct strategies, Str $\in \{\mbox{Str1, Str2, Str3, Str4, Str5}\}$, and for three couples $(r,D) \in \{(0.3,0.1),(0.3,0.2),(0.8,0.1)\}$  (Fig. \ref{Fig3}). The first management strategy, Str1, splits the host population into equal proportions of a maximum number of types, $2^N$ (Fig. \ref{Fig3}A). In other words, the number of hosts receiving the multicomponent treatment is reduced to $1/2^N$, and they are mixed with untreated hosts and hosts receiving monocomponent and intermediate treatments. It should be recalled that an intermediate treatment includes only some of the available $N$ treatment components. We then progressively modified the conditions under which the multicomponent treatment is used in the management strategy. First of all, in Str2 we increased the proportion of hosts treated with multicomponent treatment to $1/2$ of the host population. Next, in Str3, we assumed that monocomponent treatments had already been overcome by the pathogen. In Str4, we eliminated intermediate treatments in order to analyze their effect on the pathogen adaptation rate. Finally, in Str5, we left only untreated hosts and hosts receiving a multicomponent treatment in equal proportions. We fixed the other parameter values as follows: the mutation cost $C=0.1$, the maximum pathogen population size $K=10000$, and the mutation rate $\nu=10^{-3}$. The initial population size of the resident pathogen on the untreated hosts was set to $\xi_1 K$ for all strategies. In Str3 and Str4, the population size of already-adapted pathogen types was set to $\xi_i K$, where $i$ is the type of host that has received a defeated treatment.\\  
  
{\bf The role of the mutation cost.}\\  
In the last part we studied the effect of the mutation cost $C \in \{0,0.1,0.2,0.3,0.4,0.5,0.6,0.8,0.9\}$ on the emergence time at the following parameter values: the number of component treatments $N=1$, the proportion of hosts treated, $\xi \in \{0.2,0.5,0.8\}$, the growth rate, $r \in \{0.3,12\}$, the migration rate, $D \in \{0.05,0.15,0.25\}$, the maximum pathogen population size $K=10000$, and the mutation rate $\nu=10^{-5}$ (not illustrated). The mutation cost impact $C \in \{0,0.1,0.2,\}$  was also studied at the number of treatment components $N \in \{2,3\}$ for the treatment strategy Str $\in \{\mbox{Str2, Str5}\}$, the growth rate, $r =0.3$, the migration rate, $D \in \{0.05,0.25\}$, the maximum pathogen population size $K=10000$, the mutation rate $\nu=10^{-3}$, and the initial population size of the resident pathogen type on the untreated hosts was set to $\xi K$ (Fig. \ref{Fig2}). \\  

{\bf Model implementation.}\\  
  We applied the Gillespie algorithm \cite{Gil1977} to track the exact trajectories of the population size of pathogen types, $X_{ij}(t)$, driven by the birth and death process. Since the algorithm is computationally expensive for large populations, we used a Gillespie method up to a certain population size and then, when the population size of any pathogen type reached high values and approached its equilibrium, we used its deterministic equilibrium calculated from the corresponding system of differential equations:
\begin{equation}\label{eqdiff}
\frac{dX_{ij}(t)}{dt}=\lambda_{ij}+\Lambda_{ij} - \mu_{ij} - \sum \limits_{k \ne j} \gamma_{ijk} + \sum \limits_{l \ne j} \gamma_{ilj},\; \mbox{for all } i,j \in [1,2^N]^2.
\end{equation}
Indeed, it has been shown that a birth and death process converges to a continuous diffusion process when the population size is high \cite{Eth2005}. To test our algorithm, we compared it to some exact trajectories using Student's t-test. The model was implemented in C++ using Code Blocks and GNU GCC compiler.

\section*{Results} \label{res}

\subsection*{Emergence time as a U-shaped function of the proportion of the host treated }

Numerical simulations show that in the context of a two-type host population including monocomponent, treated and untreated individuals, the emergence time is a U-shaped function of the proportion of the hosts treated, $\xi$ (Fig. \ref{Fig1}A). We see that at the fixed value of $r$, the highest values of the emergence time function can correspond to either low or high $\xi$ values (Fig. \ref{Fig1}A). The impact of the proportion of the host treated on the emergence time depends on both the migration, $D$, and growth, $r$, rates of the pathogen population. Figure \ref{Fig1}B summarizes the responses of the emergence time to the variation of $r$, $D$ and $\xi$. At most combinations of $r$ and $D$, the emergence time is longer when either low, $\xi<0.3$, or high, $\xi>0.7$ proportions of the hosts were treated. However, when $D$ has a medium or high value, $D>0.1$, and $r$ are low, $r<0.5$, the emergence of an escape mutant can be impeded only by treating high proportions of the host. For low $D< 0.1$, and high $r>1$ values, small proportions provide a better control of emergence, whereas for high $D>0.2$, coupled with high $r >1$, at any proportion the emergence time is short and we cannot therefore impede rapid pathogen adaptation. If treated hosts are subjected to a multicomponent treatment, the emergence time keeps its U-shape versus $\xi$, but an increase in the component number prolongs the emergence time. In contrast to the impact of migration and growth rates on the emergence time, the maximum pathogen population size, $K$, affects only the values of the emergence time function, but not its shape. An increase in $K$ speeds up the pathogen adaptation.\\

Our model also demonstrates how mutation and migration processes drive the emergence time (Fig. \ref{Fig1}C). The proportion of hosts treated, $\xi$, determines the importance of each process in pathogen adaptation. When $\xi$ is small, the time to mutation is short and the time to migration is long. As $\xi$ increases, the mutation time grows, while the migration time decreases. If small proportions of the hosts are treated, the emergence time is the sum of the mutation time and the migration time, whereas for high proportions the emergence time is greater than this sum.

\subsection*{Management strategies and the durability of the multicomponent treatment}

We then varied the treatment strategy and the values of pathogen life-cycle parameters to three pairs of the growth and migration rates and four different numbers of treatment components $N \in \{1,2,3,4\}$. To investigate the impact of the management strategy on the durability of the multicomponent treatment, we considered five distinct management strategies, $\{\mbox{Str1, Str2, Str3, Str4, Str5}\}$, that split the host population into different proportions of various types, $\xi_j$ (Fig. \ref{Fig3}A). The strategies were designed to illustrate the effect of intermediate treatments and the presence of already defeated treatments on the durability of $N$-component treatment. We assume that this treatment can either be deployed alone or used simultaneously with other treatments involving various combinations of the same components (Fig. \ref{Fig3}A). Thus, a management strategy can deploy a $N$-component treatment with monocomponent treatments, $N=1$, and various intermediate treatments with combinations of $2$ to $N-1$ components. Figure \ref{Fig3} shows that the emergence time increases with the number of components in a treatment. Interestingly, the deployment mode of a multicomponent treatment also has a major impact on the rate of pathogen adaptation. As the proportion of the hosts carrying $N$-component treatment increases, the emergence time of an escape mutant increases progressively from Str1 to Str2. Nevertheless, the emergence time decreases if the pathogen has already adapted to hosts receiving monocomponent treatments, as in Str3. When there are no hosts receiving intermediate treatments, Str4, the emergence time increases if the number of components is equal to or greater than 3. Treatment strategy Str5, which only divides the host population into two types: untreated hosts and treated hosts receiving multicomponent treatment, is the most durable. The increase in the growth rate raises the probability of mutation, and thus shortens the emergence time in any strategy, especially if the pathogen has accumulated several mutations on the same hosts (Fig. \ref{Fig3}D). Figure \ref{Fig3}C shows that the migration rate has a more complicated impact on the emergence time than the growth rate: the increase in the migration rate prolongs the emergence time if the degree of host diversification is high, as in strategies Str1, Str2 and Str3.\\ 

\subsection*{Impact of the mutation cost on the emergence time}

We finally varied the mutation cost value for two management strategies, Str2 and Str5, two migration rates and two numbers of component treatments (Fig. \ref{Fig2}). The model shows that an increase in the mutation cost increases the time to emergence, but that the intensity of the impact of the mutation cost depends on the migration rate (Fig. \ref{Fig2}A). Indeed, the higher migration rate, the slower the increase in the emergence time with the increase in mutation cost. It is easy to see from Fig. \ref{Fig2}A that the effect of an increase in the migration rate becomes more pronounced when the number of treatment components increases from $N=2$ to $3$. For a monocomponent treatment, $N=1$, the mutation cost has almost no effect on the emergence time at high values of the migration rate (not illustrated). However, this can be altered by the deployment mode of the multicomponent treatment. Figure \ref{Fig2}B demonstrates that we can observe a delay in the emergence time at high migration rates, if the management strategy divides the host population in many types, e.g. as Str2 (Fig. \ref{Fig3}A). As before, the effect of an increasing migration rate intensifies with the number of treatment components.

\section*{Discussion}

In this article, we have developed a stochastic framework for estimating the speed of pathogen adaptation in response to a population-wide management strategy deploying a multicomponent treatment in a host population, such as  combination therapies, combination vaccines/chemicals and cultivars carrying multiple resistance genes. Our model provides a basic understanding of how life-cycle parameters of the pathogen population and controllable parameters of a management strategy affect the speed of pathogen adaptation to a multicomponent treatment. Our results reveal the importance of coupling stochastic mutation and migration processes, and illustrate how their stochasticity can alter our view of the principles of the management of the pathogen adaptive dynamics at the population level.\\

Multi-type birth and death processes are a powerful tool for modelling adaptive pathogen dynamics, since they can easily be adapted to many biological situations by adjusting the transition rates \cite{All2003,Nov2006}. This approach makes it possible to monitor the stochastic dynamics of small populations, such as  an escape mutant. Moreover, the approach makes it possible to derive an analytical estimation of the emergence time \cite{Bou2011} to study the treatment durability. Note that the model results can be affected by transition functions. However, a comparative analysis of the impact of various transition functions on the model dynamics was beyond the scope of our study. Nevertheless, our choice of transition functions allowed us to track the pathogen adaptive dynamics using just a small number of parameters. Moreover, parameters can easily be assessed from epidemiological data on the pathogen dynamics, for instance the dynamics of its population size, pathogen dispersal, and using information about the treatment used, such as the probability that a pathogen will mutate in order to escape the treatment, and the mutation cost per mutation. Note that mutation probability and pathogen population size, should be estimated most precisely, since the model is sensitive to their variations (see Supporting Information S1). It is also known that the pattern of a spatially heterogeneous treatment can have an impact on the pathogen adaptive dynamics \cite{Bon1997,Deb2009,Sap2009}. Since our model is non-spatial, it can only be applied when a treatment-diversified host population is well-mixed, and its spatial structure can be ignored. To preserve the simplicity of the model and the coherence of the results, we did not consider either compensatory mutations or recombinations that could accelerate the emergence of mutants escaping a multicomponent treatment \cite{Han2006,Rho2005,Mac2010,Day2012}. 

\subsection*{The proportion of the host treated and the speed of pathogen adaptation}

Our model shows that the optimum proportion of hosts treated in order to impede the adaptation of the pathogen population depends on the interplay between the intensity of pathogen reproduction and the migration processes. If both processes have high rates, then durable control is impossible, and the treatment will soon be overcome. This finding is consistent with empirical results showing that plant pathogens, such as rust and mildew, which have a high gene flow and large population size, have a long history of defeating major resistance genes and their pyramids \cite{McD2002}. If the migration rate prevails over the pathogen growth rate, then the proportion of hosts treated should be high enough to reduce the size of the resident pathogen population on the untreated hosts, and thus to reduce the probability of mutation. Conversely, when growth rate dominates over migration rate, the proportion of the host treated should be low in order to reduce the probability of migration. Overall, the proportion of hosts treated should be adjusted to control the recessive process: mutation or migration. If the two processes are equivalent, the rate of pathogen adaptation to treatment can be slowed down by treating either small or high proportions of the host. This conclusion is the same as that reached from  the fundamentally different model of  van den Bosch and Gilligan \cite{Bos2003}, which is deterministic and it does not take the mutation cost into account. However, the extension of our model to a multicomponent treatment showed that the result can be generalised: when the number of treatment components increases, the emergence time function keeps its U-shaped form. In the context of adaptation to a multicomponent treatment, an individual pathogen has to cope with more than one mutation, which makes the emergence of an escape mutant on untreated hosts a rare event, especially when the size of the untreated host subpopulation is small. Because of the additive mutation cost, mutants do not live for long on the untreated hosts, which reduces the probability of the successive migrations onto the treated host, especially when the frequency of treated hosts is low. \\

To be sustainable, a treatment strategy has to control not only the spread of the epidemic, but also the adaptive dynamics of the pathogen. Theoretical studies of epidemiology and biological invasions focusing on the control of population spread demonstrate that there is a lower limit for the proportion of treated hosts that can minimize this population expansion, which is claimed to be about seventy percent \cite{Col2000,Ott2004,Oht2006,Suz2011}. Our results suggest that this proportion does not offer a durable control strategy for a pathogen population with high reproduction and low migration rates, since it would simply accelerate its adaptation. Epidemiological models that ignore pathogen migration and the stochasticity of pathogen dynamics show that the optimal vaccination coverage that prevents the emergence of a drug resistant pathogen strain is about $1-1/R_0$  \cite{Sch2002}, where $R_0$ is the basic reproductive number. Note that this estimation includes intermediate proportions of treated hosts that, according to our results, can accelerate pathogen adaptation and thereby reduce the vaccine life span. In plant epidemiology,  Ohtsuki and Sasaki \cite{Oht2006} concluded that if there is a high risk of the development of virulent pathogen that can infect the resistant host, the fraction of resistant crop should never exceed about twenty-five percent for any pathogen having $R_0>1$. Linking the characteristics of the plant resistance level with the epidemic dynamics, Fabre et al. \cite{Fab2012} have shown that low cropping ratios of a resistant plant can prolong its durability. However, our model suggests that small fractions of the resistant crop can speed up pathogen adaptation if the pathogen population with the low growth rate is easily able to migrate. Thus, we can conclude that accounting for the stochastic migration process in modelling pathogen adaptive dynamics alters the extant criteria for the critical proportion of treated hosts that could impede pathogen adaptation. Superimposing our results over epidemiology criteria suggests the conditions required for the optimum proportion of the treated hosts leading to the control of both the evolutionary and invasive dynamics of epidemics. \\

\subsection*{Deployment of a multicomponent treatment in a population-wide management strategy}

It is commonly thought that the number of components in a treatment has a significant impact on the speed of pathogen adaptation \cite{Ham1986,And1988,Hoo2000,Kel2003}. Our model shows that in fact the adaptation speed depends on a population-wide management strategy of the multicomponent treatment, i.e. on the proportion of hosts treated and on the presence of hosts receiving intermediate treatment. Intermediate treatments, including only some of the components, make it possible for an escape mutant to establish an abundant population and to accumulate the  number of mutations required to overcome a multicomponent treatment. Our results demonstrate that the deployment of a purely multicomponent treatment is the most effective strategy for impeding pathogen adaptation, but if we cannot guarantee that no hosts receive intermediate treatments, it is better to deploy treatment components separately, thereby diversifying the host population.\\

Our model suggests that if a multicomponent treatment is deployed simultaneously with treatments that have already been overcome, the adaptation rate increases greatly. Indeed, fewer mutations are needed to create an escape mutant able to invade hosts receiving the multicomponent treatment. However, deploying treatments involving three or more defeated components and their combinations can still be effective, on condition that they have been overcome independently, and if no hosts receive intermediate treatments. The absence of intermediate treatments slows down pathogen adaptation, since the pathogen has to accumulate all the necessary mutations and then migrate successfully, whereas if some hosts receive the defeated monocomponent treatment, this limits the emergence of an escape mutant. \\

Our results explain the empirical observations of the stimulation of rapid pathogen adaptation as a result of using intermediate treatments. Despite the fact that our results correspond to instantaneous host diversification in response to treatment, we can draw an analogy with situations in which various treatments are deployed successively, thus diversifying the host population over time. For instance, to control {\it Bremia lactucae}, breeders add a new resistance gene to the lettuce cultivar after each rather rapid breakdown of lettuce resistance by the pathogen \cite{Mic2008}. In other words, a new lettuce cultivar carrying several defeated resistance genes and one undefeated resistance gene is deployed just after a cultivar has lost its resistance. This can be viewed as corresponding to an intermediate treatment, and it allows the pathogen to overcome a new resistant cultivar by just a single mutation. Since our model links pathogen adaptive dynamics to controllable parameters of treatment strategies, it can be applied to the design of sustainable strategies for the selection and deployment of new multigene resistant plants, even if they carry defeated resistance genes. 

\subsection*{Mutation cost and treatment durability}

Our model confirms the fact that the mutation cost increases the emergence time \cite{Car2001,Ber2006,Fab2009,Mac2010}, and that it is not the only parameter controlling the emergence of an escape mutant \cite{Lan1997}. Moreover, our model shows that a high migration rate can mitigate the impact of the mutation cost on the emergence time. Indeed, the mutation cost essentially determines the intensity of the competition between resident and mutant pathogen types on the untreated hosts. When the migration rate is high, mutants more easily escape the fate to be outcompeted by resident individuals. When the migration rate is low, the probability of escaping is low, since a mutant outcompeted by a resident type is not able to develop an abundant population to produce a migration event. Interestingly, management strategies that split up the host population into numerous types can enhance the impact of the mutation cost on the pathogen adaptation speed, even if the pathogen has a high migration rate. In the diversified host population, the number of suitable hosts for intermediate mutants is greatly reduced, and high migration rates increase the probability of migrating onto unsuitable hosts resulting in extinction, thereby increasing the emergence time. We conclude that we should abandon the generally-accepted belief that mutation cost determines treatment durability \cite{Lea2001}, since, as we have seen, the biological context can greatly alter the impact of the mutation cost on the adaptation rate. 

\subsection*{Conclusions}
In this paper we show that considering  the interactions between two stochastic evolutionary forces, mutation and migration, can increase our understanding of the adaptation process at the population level. Since our model is appropriate for diseases with both direct and vector-borne transmission, it can be applied to designing sustainable, population-wide management strategies for a large class of harmful organisms. The model can be easily extended to the description of progressive pathogen adaptation, such as declining efficacy of imperfect vaccines or the erosion of partial cultivar resistance. Moreover, due to its simple structure, the model can be modified to account for the interactions between treatment components in order to study their effects on the speed of pathogen adaptation.


\section*{Acknowledgments}
The authors would like to thank Fr\'ed\'eric Fabre for discussions and anonymous reviewers for helpful comments on the manuscript.

\bibliography{template}

\newpage

\section*{Figure Legends}

\begin{figure}[h]
\begin{center}
\includegraphics[height=16cm]{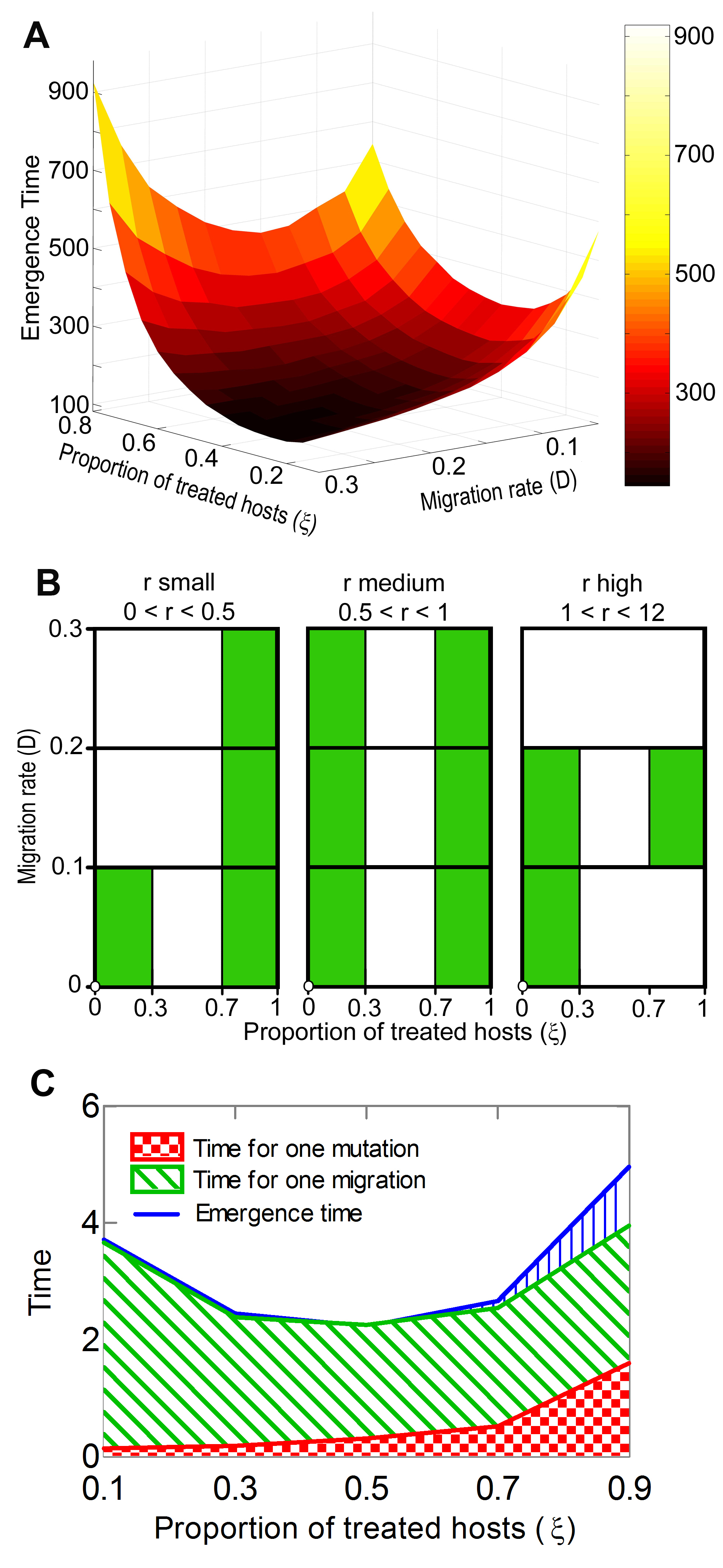}
\end{center}
\label{Fig1b}
\end{figure}


\begin{figure}[p]
\begin{center}
\end{center}
\caption{
{\bf Model predictions of emergence time dynamics.} We assume a management strategy that divides the host population into untreated and monocomponent treated individuals, so that the proportion of treated hosts is $\xi$. Here, the emergence time of an escape mutant is the time before it migrates from an untreated into a treated host. (A) Emergence time as a function of the migration rate, $D$, and the proportion of hosts treated, $\xi$. The results are based on a simulation of model equations (\ref{EqB}-\ref{EqITime}) with the following parameter values: $r=0.3$, $\nu = 10^{-5}$, $C = 0.2$, $K=10000$. (B) A simplified decision diagram to assist with developing management strategies in order to achieve durable pathogen control. The decision diagram sums up the emergence time functions obtained as in (A), but with a tuning growth rate, $r$, from 0 to 12 and $D$ from $0.1$ to 0.3. All the other parameter values are identical to those in Figure \ref{Fig1}A. For any pair of pathogen parameters ($r$, $D$), the diagram depicts in green the proportions of treated hosts, $\xi$, that could inhibit pathogen adaptation. For instance, a pathogen with high growth and dispersal rates adapts swiftly at any proportion of the host treated, while the adaptation of a pathogen with intermediate growth and migration rates can be inhibited by either low or high proportions. (C) Emergence time decomposition. Mean times of mutation, mutant migration, and mutant emergence on a treated host as functions of the proportion of the host treated, $\xi$. The mutation time is the time the first pathogen mutation occurs on an untreated host. The migration time is the time of the first migration of the pathogen mutant from an untreated to a treated host. To reduce the calculation time, simulations were performed with the following parameter values: $r = 0.8$, $D=0.2$, $\nu = 10^{-3}$, $C=0.1$, $N=1$, $K=10000$. In all simulations the initial population size of the resident pathogen type is $X_{11}(0)=(1-\xi)K$ and for all other types $(i,j) \ne (1,1)$, $X_{ij}(0)=0$.
}
\label{Fig1}
\end{figure}

\newpage

\begin{figure}[h]
\begin{center}
\includegraphics[height=16cm]{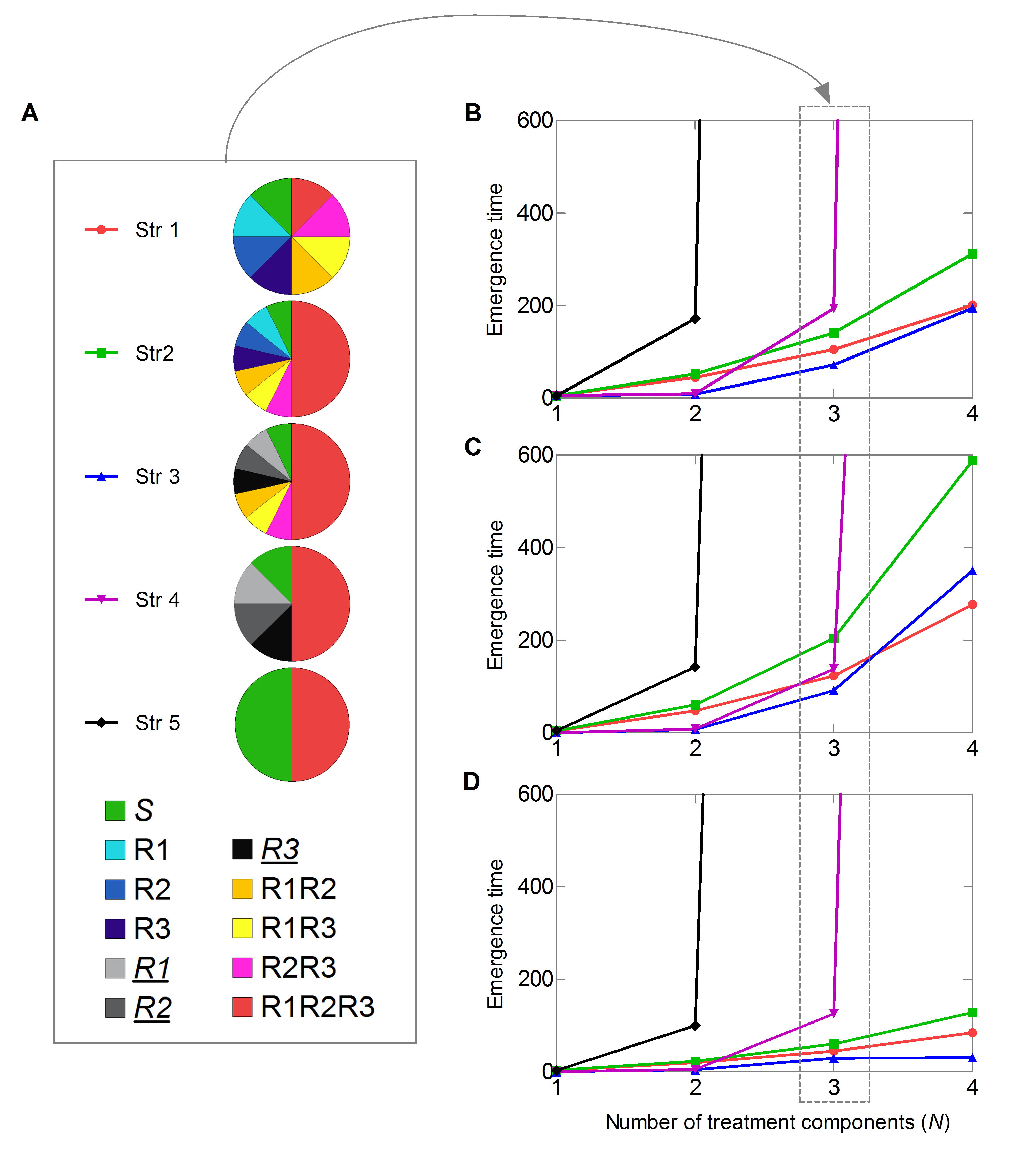}
\end{center}
\label{Fig3b}
\end{figure}


\begin{figure}[p]
\begin{center}
\end{center}
\caption{
{\bf Emergence time as a function of the number of treatment components ($N$)} for five different management strategies, Str1, $\dots$, Str5. Pie charts (A) depict the composition of a diversified host population by the deployment of a three-component treatment ($N=3$): each color corresponds to a host type listed below, and the proportion of each host type corresponds to a pie sector. Host type \textit{S} includes untreated hosts, types {R1, R2, R3} and {\underline{\textit{R1}}, \underline{\textit{R2}}, \underline{\textit{R3}}} - hosts receiving different monocomponent treatments, host types {R1R2, R1R3, R2R3} receiving different intermediate treatments, and type R1R2R3 consists of hosts receiving a three-component treatment. Host types with \underline{{\it italic and underline}} names receive defeated treatments that have already been overcome. For the detailed description of the management strategies, see the text in Materials and Methods section. We used the same deployment principle for the $N$-component treatment, $N \in \{1,\dots,4\}$. In (B), the pathogen has low growth and migration rates,  $r=0.3$ and $D=0.1$. Emergence time takes high values for $\mbox{Str}=4$ and $N=4$, $S_{4,4}=13764.3$, and for $\mbox{Str}=5$ and $N=3$, $S_{5,3}=12372.7$. In (C), the pathogen has a low growth rate and a medium migration rate, $r=0.3$ and $D=0.2$. $S_{4,4}=4721.4$ and $S_{5,3}=9440.18$ are the highest values of the emergence time. In (D), the pathogen has a medium growth rate and a low migration rate, $r=0.8$ and $D=0.1$. $S_{4,4}=9863.58$ and $S_{5,3}=8451.13$ are the highest values of the emergence time. In (A-C), the other parameters are $\nu = 10^{-3}$, $K=10000$ and $ \; C=0.1$. In Str1, Str2 and Str5, the initial population size of the resident pathogen type is $X_{11}(0)=(1-\xi)K$ and for all other types $(i,j) \ne (1,1)$, $X_{ij}(0)=0$. In Str3 and Str4, $X_{11}(0)=(1-\xi)K$, $X_{kk}(0)=\xi_i K$ where $k$ corresponds to a host type treated with a monocomponent treatment, and for all other types $(i,j) \ne (1,1)$ and $(i,j) \ne (k,k)$ , $X_{ij}(0)=0$. }

\label{Fig3}
\end{figure}




\begin{figure}[p]
\begin{center}
\includegraphics[]{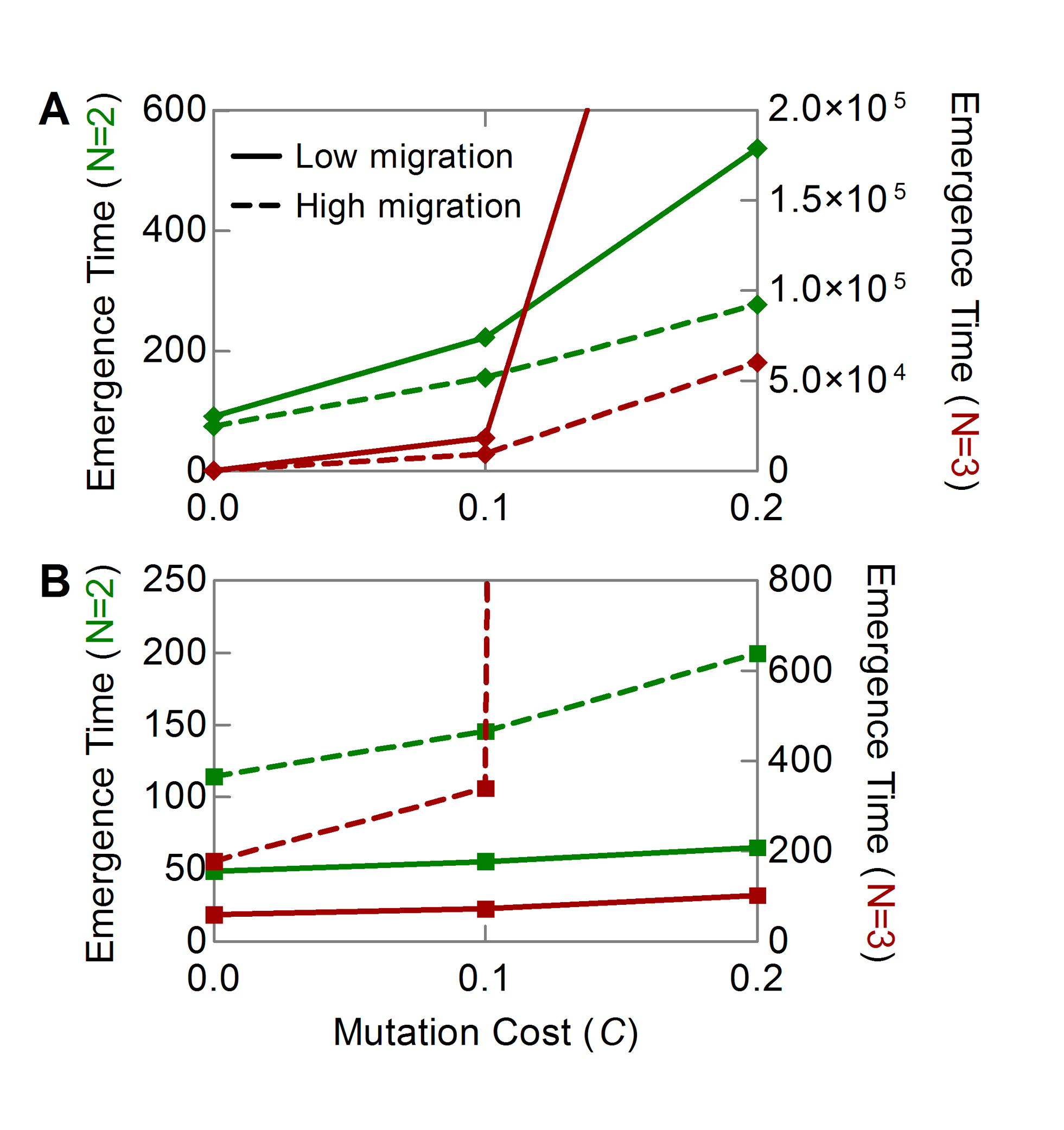}
\end{center}
\caption{
{\bf Emergence time as a function of the mutation cost.} The emergence time is plotted for a two-component treatment ($N=2$, green, left axis) and a three-component treatment ($N=3$, red, right axis) at low (solid line) and high (dotted line) migration rates. (A) The management strategy splits the host population into a 1:1 mixture of untreated and multicomponent-treatment receiving hosts (Str5, Fig. \ref{Fig3}A). (B) Multicomponent treatment is deployed with other treatments involving different combinations of the $N-1$ components, so that the host type carrying the muticomponent treatment constitutes 50\% of the host population, and the other host types are present in equal proportions (Str2, Fig. \ref{Fig3}A). In (A) and (B), we let $r = 0.3$, $D_{low}=0.05$, $D_{high}=0.25$, $\nu = 10^{-3}$ and $K=10000$. In all simulations the initial population size of the resident pathogen type is $X_{11}(0)=(1-\xi)K$, and for all $(i,j) \ne (1,1)$, $X_{ij}(0)=0$.
}
\label{Fig2}
\end{figure}

\newpage

\section*{Tables}

\renewcommand{\baselinestretch}{1.5}
\begin{table}[h]

\caption{\bf{Definition of variables and parameters used to model pathogen adaptive dynamics.}}
\begin{tabular}{|c|c|p{9cm}|c|}
\hline
Name & Value & Description& Reference\\
\hline
$X_{ij}(t)$ & - & The random variable for the number of pathogen individuals of type $i$ on host type $j$ at time $t$&-\\
\hline
$X(t)$ & $(X_{ij}(t))_{i,j \in [1,2^N]^2}$ & Vector random variable for the number of each type in the pathogen population at time $t$&-\\
\hline
$r$ & $[0.3;12]$ & Growth rate of pathogen population&\cite{Cue2005,Fab2009}\\
\hline
$D$ & $[0.1;0.3]$ & Migration rate of pathogen population&-\\
\hline
$D_{ik}$& $0$ or $D$ &$D_{ik}=D$ if pathogen type $i$ can infect host type $k$, else $D_{ik}=0$&-\\
\hline
$N$ & $\{1, 2, 3, 4\}$ & Number of treatment components&-\\
\hline
$C$ & $[0;0.8]$ & Mutation cost of carrying one mutation&\cite{Aye2006,Car2007,Lab2009}\\
\hline
$\beta_i$ & $1-\sum_{z \in i} C$ & Fractional reduction due to mutation cost of the growth rate of pathogen type $i$ & -\\
\hline
{$p_{ki}$} & $0$ or $\nu$, $\nu \in [10^{-5};10^{-3}]$& {Probability that a pathogen type $k$ will to mutate into a pathogen type $i$}& \cite{Kim1968,Fab2009,San2010}\\
\hline
$p_i$ & $\sum_{l \ne i} p_{il}$  &  Mutation probability for pathogen type $i$&-\\
\hline
$\xi_j$ & $[0;1]$ & Proportion of host type $j$ in the host population&-\\
\hline
$\xi$ & $[0;1]$ & Proportion of the host treated when $N=1$&-\\
\hline
$K$ & 10000& Maximum population size of the pathogen population&-\\
\hline
$S$  & Equation (\ref{defS}) & Emergence time of an escape mutant able to infect an individual host carrying $N$-component treatment&- \\
\hline
\end{tabular}

\label{param}
\end{table}

\newpage

\section*{Supporting Information Legends}

\textbf{- Supporting Information S1: The distribution law of the emergence time S (7). \\}
The model shows that for strategies Str5 and Str4, when the number of treatment components, N, increases, the law of the emergence time approaches an exponential one. Conversely, when the number of treatment components is small, the law moves away from an exponential distribution and towards a normal one.\\
\textbf{- Supporting Information S2: A variance-based global sensitivity analysis of model (1-6).\\}
We find that all parameters and almost all the interactions between them have a significant effect on the emergence time S.

\end{document}